\documentclass[runningheads]{svmult}

\usepackage{makeidx}   
\usepackage{graphicx}  
\usepackage{subeqnar}  
\usepackage{multicol}  
\usepackage{physprbb}  
\makeindex             

\begin{document}
\title*{Star Cluster Systems in Interacting and Starburst Galaxies: A
Multicolour Approach}
\toctitle{Star Cluster Systems in Interacting and Starburst Galaxies:
\protect\newline A Multicolour Approach}
%
%
\titlerunning{Star Cluster Systems in Interacting Galaxies}
%
\author{Richard de Grijs}
\authorrunning{Richard de Grijs}
%
%
\institute{Institute of Astronomy, University of Cambridge, Madingley
Road, Cambridge CB3 0HA, UK}

\maketitle              

\def\asec {{$\buildrel{\prime\prime}\over .$}}

\begin{abstract}

The large majority of extragalactic star cluster studies done to date
have essentially used two or three-passband aperture photometry,
combined with theoretical stellar population synthesis models, to obtain
age estimates.  The accuracy to which this can be done depends on the
number of observations through different (broad-band) filters available
as well as, crucially, on the actual wavelength range covered.  I show,
based on the examples of the nearby starburst galaxies NGC 3310 and M82,
that the addition of, in particular, near-infrared passbands to a set of
optical filters greatly enhances our ability to disentangle age,
metallicity and extinction parameters for star clusters with ages
younger than a few Gyr.  In addition, for the intermediate-age star
cluster system in the fossil starburst region of M82 we find (i) a
well-defined burst of cluster formation at slightly older ages than
derived from previous estimates based on optical fluxes alone, and (ii)
that by considering {\it only} the clusters originating in this burst of
cluster formation, we uncover the first conclusive observational
evidence for an unambiguous turn-over in the luminosity function of a
coeval star cluster system with an age as young as $\sim 1$ Gyr. 

\end{abstract}

\section{The cluster luminosity function as diagnostic}

The distribution of cluster brightnesses, known as the Cluster
Luminosity Function (CLF), is one of the most important diagnostics for
the study of extragalactic compact star cluster populations.  For the
old globular cluster (GC) systems the CLF shape is well-established: it
is roughly Gaussian, with the peak or turn-over magnitude at $M_V^0
\simeq -7.4$ and a FWHM of $\sim 3$ mag (Harris 1991, Whitmore et al. 
1995, Harris et al.  1998).  The well-studied young star cluster (YSC)
population in the Large Magellanic Cloud, on the other hand, displays a
power-law CLF (Elson \& Fall 1985, Elmegreen \& Efremov 1997). 

{\sl Hubble Space Telescope (HST)} observations have provided CLFs for
young compact cluster systems in more distant galaxies and are
continuing to do so.  Although incompleteness effects often preclude
detection of a turn-over in the CLF (e.g., Whitmore \& Schweizer 1995,
Schweizer et al.  1996, Miller et al.  1997), in galaxies with YSC
systems for which deep observations are available there is little
evidence for an intrinsic turn-over (but see de Grijs, Bastian \& Lamers
2002b).  In most of these cases, the CLF shapes are consistent with
power laws down to the completeness threshold (but see Miller et al. 
1997, de Grijs et al.  2001, 2002b). 

GC formation models suggest that the distribution of the initial cluster
masses (and, therefore, of the initial cluster luminosities) is closely
approximated by a power law (e.g., Harris \& Pudritz 1994, McLaughlin \&
Pudritz 1996, Elmegreen \& Efremov 1997).  The processes responsible for
the depletion of, preferentially, low-luminosity, low-mass star clusters
over time-scales of a Hubble time, leading to the Gaussian CLFs
observed, include tidal interactions with the gravitational field of the
parent galaxy and evaporation of stars through two-body relaxation
within clusters.  From the models of Gnedin \& Ostriker (1997) and
Elmegreen \& Efremov (1997) it follows that {\it any} initial mass (or
luminosity) distribution will shortly be transformed into peaked
distributions.  However, Vesperini (2000, 2001) has demonstrated that,
due to dynamical friction effects affecting the high-mass clusters,
considerable fine tuning of the model parameters is required to produce
from an initial power-law distribution a Gaussian-type mass function
with parameters similar to those observed for well-studied GC systems. 

However, all of these models are valid {\it only} for time-independent
Milky Way-type gravitational potentials; galaxy-galaxy interactions will
obviously have a major effect on the resulting (time varying)
gravitational potential, in which the dynamical star cluster evolution
is likely significantly different. 

\section{The importance of multicolour photometry}

Age spread effects in cluster systems in which cluster formation is
still ongoing affect the observed CLF (Meurer 1995, Fritze--v. 
Alvensleben 1999, de Grijs et al.  2001, 2002a,b), which might in fact
make an intrinsically Gaussian CLF appear as a power-law CLF (see, e.g.,
Miller et al.  1997).  For young and intermediate-age cluster systems it
is very important to age date the individual clusters and to correct the
observational CLF to a common age before interpreting their CLF. 

Metallicities of YSCs produced in galaxy interactions, mergers or
starbursts are an important discriminator against GCs formed in the
early Universe.  They are expected to correspond to the interstellar
medium abundances of the interacting/starburst galaxies and are
therefore most likely significantly higher than those of halo GCs in the
Milky Way and other galaxies with old GC systems.  Precise metallicity
determinations for individual YSCs are therefore important to determine
ages, and age spreads, from integrated colours. 

Dust extinction is often very important in YSC systems.  In particular
the youngest post-burst galaxies and galaxies with ongoing starbursts
often show strong and patchy dust structures and morphologies. 
Extinction estimates towards individual YSCs are therefore as important
as individual metallicity estimates in order to obtain reliable ages and
to be able to derive an age-normalised CLF or YSC mass function. 

As part of an ongoing ASTROVIRTEL\footnote{The ASTROVIRTEL project is
aimed at enhancing the scientific return of the ST-ECF/ESO Archive.  It
offers the possibility to European users to exploit it as a virtual
telescope with the assistance of the Archive management.} project, we
are in the process of assessing the systematic uncertainties in age,
extinction and metallicity determinations for YSC systems inherent to
the use of broad-band, integrated colours.  We have developed an
evolutionary synthesis optimisation technique that can be applied to
photometric measurements in a given number $N (N \ge 4)$ of broad-band
passbands (see Anders, Fritze--v.  Alvensleben \& de Grijs, these
proceedings).  The optimisation routine then simultaneously determines
the best combination of age, extinction and metallicity from a
comparison with the most up-to-date G\"ottingen simple stellar
population (SSP) models (Schulz et al.  2002), to which we have added
the contributions of an exhaustive set of gaseous emission lines and
gaseous continuum emission. 

The most important scientific issue we want to address is whether, and
to what accuracy, we can determine the violent star formation histories
of galaxies based on star cluster colour distributions and deep
luminosity functions in several optical and/or near-infrared (NIR)
passbands.  Closely related to this issue is the question as to whether
the star clusters formed in interacting galaxies will eventually evolve
into the old GC systems we see today.  The large majority of
extragalactic star cluster studies done to date have essentially used
two or three-passband aperture photometry, combined with theoretical
stellar population synthesis models, to obtain age estimates (while in
most cases metallicities were fixed at solar values).  The accuracy to
which this can be done obviously depends on the number of observations
through different (broad-band) filters available as well as, crucially,
on the actual wavelength range covered. 

\section{A case for additional near-infrared observations}

As a pilot case for our study of systematic effects introduced by the
choice of any particular passband combination we decided to concentrate
on NGC 3310.  This choice was made because of (i) the large and
homogenous set of observations of this galaxy available in the {\sl HST}
archive, and (ii) the large number of prior extragalactic star cluster
studies available in the literature to which we can compare our results. 

NGC 3310 is a local ($D = 12.5$ Mpc), very active starburst galaxy with
high global star formation efficiency.  A number of morphological and
kinematical peculiarities suggest that NGC 3310 was affected by a major
gravitational disturbance, which led to high, possibly sustained, star
formation rates in the past $\sim 100$ Myr (cf.  Balick \& Heckman
1981).  The bar-driven star formation scenario suggested by Conselice et
al.  (2000), combined with the recent infall of a companion galaxy, is
the currently most attractive scenario for this: it provides a natural
explanation for the low metallicity observed in the star-forming knots
near the centre, while it also explains why we observe concentrated star
formation in star clusters or very luminous H{\sc ii} regions in a
tightly-wound ring-like structure surrounding the centre.  All of the
observational evidence points at very recent star formation in the star
clusters and H{\sc ii} regions, and a time since the interaction of $\le
10^{7-8}$ yr (e.g., Balick \& Heckman 1981, Smith et al.  1996). 

\begin{figure}[t]
\begin{center}
\includegraphics[width=1.03\textwidth]{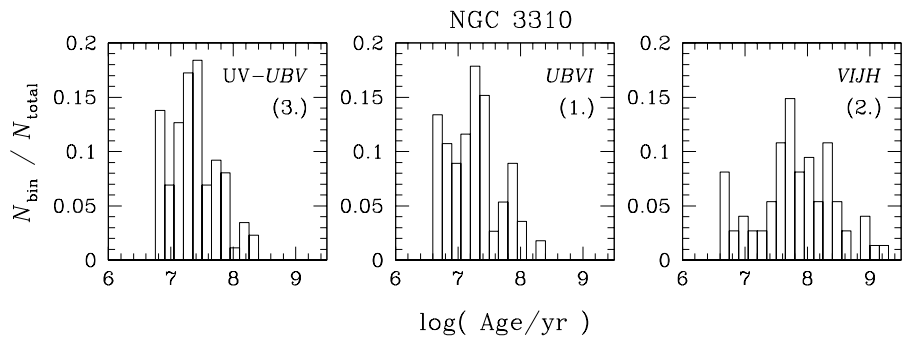}
\end{center}
\vspace{-8cm}
\caption[]{Normalised age distributions of the NGC 3310 clusters based
on three choices of broad-band passband combinations, as described in
the text.}
\label{fig1} 
\end{figure}

In the galaxy's central 22\asec55 $\times$ 22\asec65 region covered by
all of the {\sl WFPC2} {\it and} NICMOS observations available in the
{\sl HST} archive, we detected 289 point-like (or marginally extended)
star cluster candidates brighter than $4 \times$ the r.m.s.  noise of
the background in both of the F606W and F814W source lists.  To assess
the effects of choosing a particular passband combination on the final
results, we applied our fitting technique to our flux measurements in
four sets of passbands (see Fig.  \ref{fig1}):

\begin{enumerate}

\item a subset of only optical passbands: F336W, F439W, F606W and F814W;

\item a red-selected passband combination: F606W, F814W, F110W and
F160W;

\item a blue-selected passband combination: F300W, F336W, F439W and
F606W;

\item the full set of seven passbands, from F300W to F160W.

\end{enumerate}

It is immediately clear from Fig.  \ref{fig1} that the resulting age
distribution of an extragalactic star cluster system based on integrated
broad-band colours is a sensitive function of the passbands covered by
the observations.  The cause for these significantly different age
distributions with and without the inclusion of the near-UV and/or NIR
filters is the degeneracy between age and extinction, and to a lesser
extent also between age and metallicity, for the optical (and UV) colour
combinations.  In particular for the younger ages, the least-squares
fitting routines tend to overestimate the extinction without NIR
information, and therefore underestimate the ages.  As I will show below
(see also de Grijs et al.  2002a; Parmentier, de Grijs \& Gilmore, these
proceedings) for the star cluster population in M82, {\em the addition
of NIR passbands greatly enhances our ability to disentangle age,
metallicity and extinction parameters for YSCs with ages younger than a
few Gyr.} Although the full interpretation and details of our study of
systematic uncertainties will be described in detail in a forthcoming
paper, here I want to point out that while the actual age estimates are
a sensitive function of passband choice, the relative age distributions
resulting from all of the different passband combinations indicate a
young age peak (age $\le 10^7$ yr) and a major burst of cluster
formation between $10^7$ and $10^8$ yr ago.  These age estimates are
fully consistent with previously published estimates, sometimes based on
significantly more sophisticated techniques.  In fact, while the
clusters in the older burst are smoothly distributed throughout the
centre of the galaxy and its nuclear ring, the youngest star clusters
are predominantly located in the (very young) ``Jumbo'' region and the
actively star forming northern spiral arm. 

\section{The M82 fossil starburst revisited}

In a recent study focusing on the fossil starburst site in the nearby,
``prototypical'' starburst galaxy M82 we found a large population of
($\sim 110$) evolved compact star clusters (de Grijs et al.  2001),
whose properties appear to be consistent with them being evolved (and
therefore faded) counterparts of the young star clusters detected in the
galaxy's active core (O'Connell et al.  1995).  Based on broad-band
optical colours {\it only} and on comparison with stellar evolutionary
synthesis models, we estimated ages for these clusters from
$\sim 30$ Myr to over 10 Gyr, with a peak near 650 Myr.  We have since
obtained new age and mass estimates for these star clusters, based on
improved fitting methods to the {\it full} spectral energy distribution,
including the F110W and F160W fluxes (de Grijs et al.  2002a).  Our new
age estimates confirm the peak in the age histogram attributed to the
last tidal encounter with M81, but at slightly older ages than
previously published, $\log( t_{\rm peak} / {\rm yr} ) = 9.04$.  {\em
This, again, underscores the importance of including NIR passbands when
trying to determine global cluster properties from integrated
photometry.}

\begin{figure}[t]
\begin{center}
\includegraphics[width=.9\textwidth]{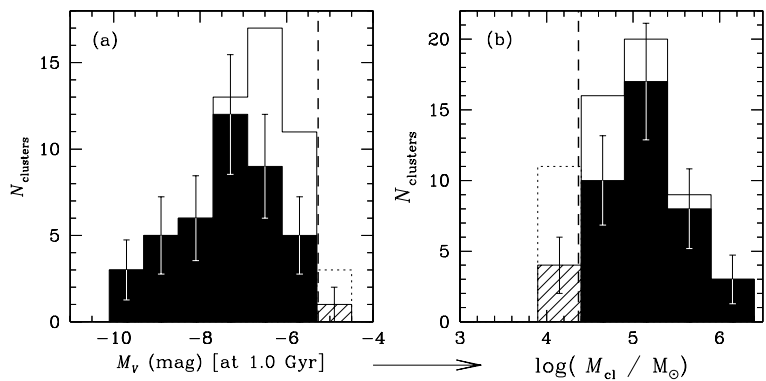}
\end{center}
\vspace{-5.5cm}
\caption[]{Age-normalised CLF and corresponding mass distribution of the
M82 clusters formed in the burst of cluster formation, $8.4 \le \log
({\rm Age/yr}) \le 9.4$.  The shaded histograms correspond to the
clusters with well-determined ages; the open histograms represent the
entire cluster sample in the age range of the burst.  The vertical
dashed line is our selection limit.}
\label{fig2} 
\end{figure}

As shown in Fig.  \ref{fig2}, both the mass distribution and the CLF
(corrected to a common age of 1.0 Gyr) of the clusters formed during the
burst, defined as $8.4 < \log({\rm Age/yr}) < 9.4$, show an unambiguous
turn-over at an order of magnitude more massive, and about 2 magnitudes
brighter, than our detection limit, respectively (de Grijs et al. 
2002b).  This turn-over is not due to selection effects, which we do, in
fact, understand very well for the clusters formed in the burst (see de
Grijs et al.  2002a).  {\em This is the first time that a clear
turn-over has been detected unambiguously for a coeval star cluster
system with an intermediate age as young as $\sim 1$ Gyr.} We have also
shown that with the very short characteristic cluster disruption
time-scale governing M82's fossil starburst region (de Grijs et al. 
2002a), its cluster mass distribution will evolve towards a higher
characteristic mass scale than for the Galactic GCs by the time it
reaches a similar age (de Grijs et al.  2002b).  We argue, therefore,
that this evidence, combined with the similar cluster sizes (de Grijs et
al.  2001), lends strong support to a scenario in which the current
slightly evolved M82 cluster population will eventually evolve into a
significantly depleted old Milky Way-type GC system dominated by a small
number of high-mass clusters.  This implies that progenitors of Milky
Way-type GCs, which were once thought to be the oldest building blocks
of galaxies, are still forming today in galaxy interactions and mergers. 
However, they will likely be more metal-rich than the present-day old GC
systems. 

\section*{Acknowledgements}

I acknowledge support from ASTROVIRTEL, a project funded by the European
Commission under FP5 Contract No.  HPRI-CT-1999-00081.  Funding for this
work was provided by the Particle Physics and Astronomy Research Council
(PPARC) in the UK. 

%

%

\end{document}